\documentclass[twocolumn,aps,prl,showpacs,superscriptaddress]{revtex4-1}
\usepackage[latin9]{inputenc}
\usepackage{amsmath}
\usepackage[dvipdfmx]{graphicx}

\makeatletter

\usepackage{verbatim}
\usepackage{color}
\usepackage{subfigure}
\usepackage{here}
\usepackage{siunitx}

\newcommand{\Isw}{I_\mathrm{sw}}

\makeatother

\begin{document}
\title{Quasiparticle trapping at vortices producing Josephson supercurrent enhancement} 

\author{Yosuke Sato}
\email{yosuke.sato@riken.jp}
\thanks{Equally contributed.}
\affiliation{Department of Applied Physics, University of Tokyo, 7-3-1 Hongo, Bunkyo-ku, Tokyo 113-8656, Japan}
\affiliation{Center for Emergent Matter Science, RIKEN, 2-1 Hirosawa, Wako-shi, Saitama 351-0198, Japan}

\author{Kento Ueda}
\email{kento.ueda@riken.jp}
\thanks{Equally contributed.}
\affiliation{Department of Applied Physics, University of Tokyo, 7-3-1 Hongo, Bunkyo-ku, Tokyo 113-8656, Japan}

\author{Yuusuke Takeshige}
\affiliation{Department of Applied Physics, University of Tokyo, 7-3-1 Hongo, Bunkyo-ku, Tokyo 113-8656, Japan}

\author{Hiroshi Kamata}
\affiliation{Laboratoire de Physique de l'\'{E}cole Normale Sup\'{e}rieure, ENS,
PSL Research University, CNRS, Sorbonne Universit\'{e}, Universit\'{e} Paris Diderot,
Sorbonne Paris Cit\'{e}, 24 rue Lhomond, 75231 Paris Cedex 05, France}

\author{Kan Li}
\affiliation{Beijing Key Laboratory of Quantum Devices, Key Laboratory
for the Physics and Chemistry of Nanodevices and Department of Electronics,
Peking University, Beijing 100871, China}

\author{Lars Samuelson}
\affiliation{Division of Solid State Physics
and NanoLund, Lund University, Box 118, SE-221 00 Lund, Sweden}

\author{H. Q. Xu}
\email{hqxu@pku.edu.cn}
\affiliation{Beijing Key Laboratory of Quantum Devices, Key Laboratory
for the Physics and Chemistry of Nanodevices and School of Electronics,
Peking University, Beijing 100871, China}
\affiliation{Division of Solid State Physics
and NanoLund, Lund University, Box 118, SE-221 00 Lund, Sweden}
\affiliation{Beijing Academy
of Quantum Information Sciences, Beijing 100193, China}

\author{Sadashige Matsuo}
\email{sadashige.matsuo@riken.jp}
\affiliation{Center for Emergent Matter Science, RIKEN, 2-1 Hirosawa, Wako-shi, Saitama 351-0198, Japan}
\affiliation{JST, PRESTO, 4-1-8 Honcho, Kawaguchi, Saitama 332-0012, Japan}

\author{Seigo Tarucha}
\email{tarucha@riken.jp}
\affiliation{Center for Emergent Matter Science, RIKEN, 2-1 Hirosawa, Wako-shi, Saitama 351-0198, Japan}
\affiliation{Department of Physics, Tokyo University of Science, 1-11-2 Fujimi, Chiyoda-ku, Tokyo 102-0071, Japan}
\begin{abstract}
  The Josephson junction of a strong spin-orbit material under a magnetic field is a promising Majorana fermion candidate.
  Supercurrent enhancement by a magnetic field has been observed in the InAs nanowire Josephson junctions and assigned to a topological transition.
  In this work we observe a similar phenomenon but discuss the non-topological origin by considering trapping of quasiparticles by vortices that penetrate the superconductor under a finite magnetic field.
  This assignment is supported by the observed hysteresis of the switching current when sweeping up and down the magnetic field.
  Our experiment shows the importance of quasiparticles in superconducting devices with a magnetic field, which can provide important insights for the design of quantum qubits using superconductors.
\end{abstract}
\maketitle
\renewcommand{\thefigure}{\arabic{figure}}
\renewcommand{\theequation}{\arabic{equation}}
\setcounter{figure}{0}

\subsection*{Introduction}
Combining an s-wave superconductor with a semiconductor nanowire (NW) made of strong spin-orbit interaction (SOI) materials, such as InAs and InSb, is of experimental interest, because it induces a topological transition to the topological superconductor (TSC) phase with suitable magnetic fields and chemical potential~\cite{alicea2012new,oreg2010helical}. 
The TSC phase of the NW coupled to the superconductor has Majorana fermions (MFs) at the edge. 
The MFs are expected to be applied to topologically protected quantum computing because of their non-abelian statistics, and recently, research on superconductor-semiconductor NW hybrid systems has been developed to find and control the MFs~\cite{sarma2006topological,leijnse2012introduction}.
In the literature, the zero-bias conductance peak~\cite{mourik2012signatures,das2012zero,deng2012anomalous,deng2014parity,albrecht2016exponential}, missing odd Shapiro steps~\cite{rokhinson2012fractional}, Josephson emission at half of the fundamental radiation frequency~\cite{laroche2019observation}, and 
enhancement of supercurrent (SC) \cite{tiira2017magnetically} have been presented as experimental evidence of MFs in the TSC phase. 
In these studies, the magnetic field is a crucial parameter that induces nontrivial topological states.
However, there have been criticisms of the experimental evidence. Critics argue that the observed phenomena can arise from a trivial source unrelated to the MFs. 
For example, the zero-bias conductance peak can be attributed to the Andreev bound state (ABS)~\cite{liu2018distinguishing,Reeg2018a,prada2020andreev,Yu2021,Valentini2021} or weak anti-localization~\cite{pikulin2012zero}. 
In addition, the missing odd integer Shapiro steps can be explained by non-adiabatic dynamics such as the Landau-Zener transition of the highly transparent Josephson junctions~\cite{dartiailh2021missing}. 
These recent criticisms indicate that thorough experimental study and careful data analysis to identify the origin of the novel superconducting transport phenomena are of great importance, significantly, for the establishment of not only TSC and MF physics but also the development of superconducting device physics.

In this report, we focus on the magnetic field-induced enhancement of SC
in the Josephson junction of a single InAs NW~\cite{tiira2017magnetically}. 
This enhancement shows that the switching current almost doubles above a certain magnetic field $B^*$ as the positive out-of-plane magnetic field is swept from \SI{0}{mT}. 
In the first place, this previous report discusses that the most critical contribution to the enhancement is the existence of MFs induced by the magnetic field. 
In contrast, we have previously reported a similar experimental study~\cite{kamata2018anomalous} on the Josephson junction of an InAs single NW, where we found an enhancement of SC for the in-plane parallel magnetic field in the NW direction. 
We concluded that this enhancement can be interpreted as low-pass filter formation, which is not related to the presence of MFs. 
However, in this experimental report, we did not study the out-of-plane SC enhancement which is the most remarkable in the previous report~\cite{tiira2017magnetically}.
Therefore, it is valuable to revisit the SC enhancement with the out-of-plane magnetic field.

For this purpose, we fabricated a Josephson junction on a an epitaxially grown InAs single NW and performed a DC measurement of the SC in a dilution refrigerator. 
Consequently, we observed the enhancement of the SC, as reported in a previous study~\cite{tiira2017magnetically,kamata2018anomalous}. 
To determine the origin of the enhancement, we measured the switching current evolution with the gate voltage and magnetic field. 
Then, we found that $B^*$ does not depend on the gate voltage, and the magnetic field dependence shows a clear hysteresis with respect to the magnetic field sweep direction. 
These results suggest that the magnetic-field-induced SC enhancement is related to the vortices penetrating the superconducting electrodes. 
Thus, we assign the enhancement origin to quasiparticles trapped in the vortex cores. 
We confirmed that the $B^*$ dependence on the applied magnetic field angle supports the quasiparticle trapping scenario.
Our results will contribute to the physics of superconducting devices and especially sort anomalous superconducting transport phenomena into trivial and nontrivial topological natures.

\begin{figure}[t]
	\centering
	\includegraphics[width=1.0\linewidth]{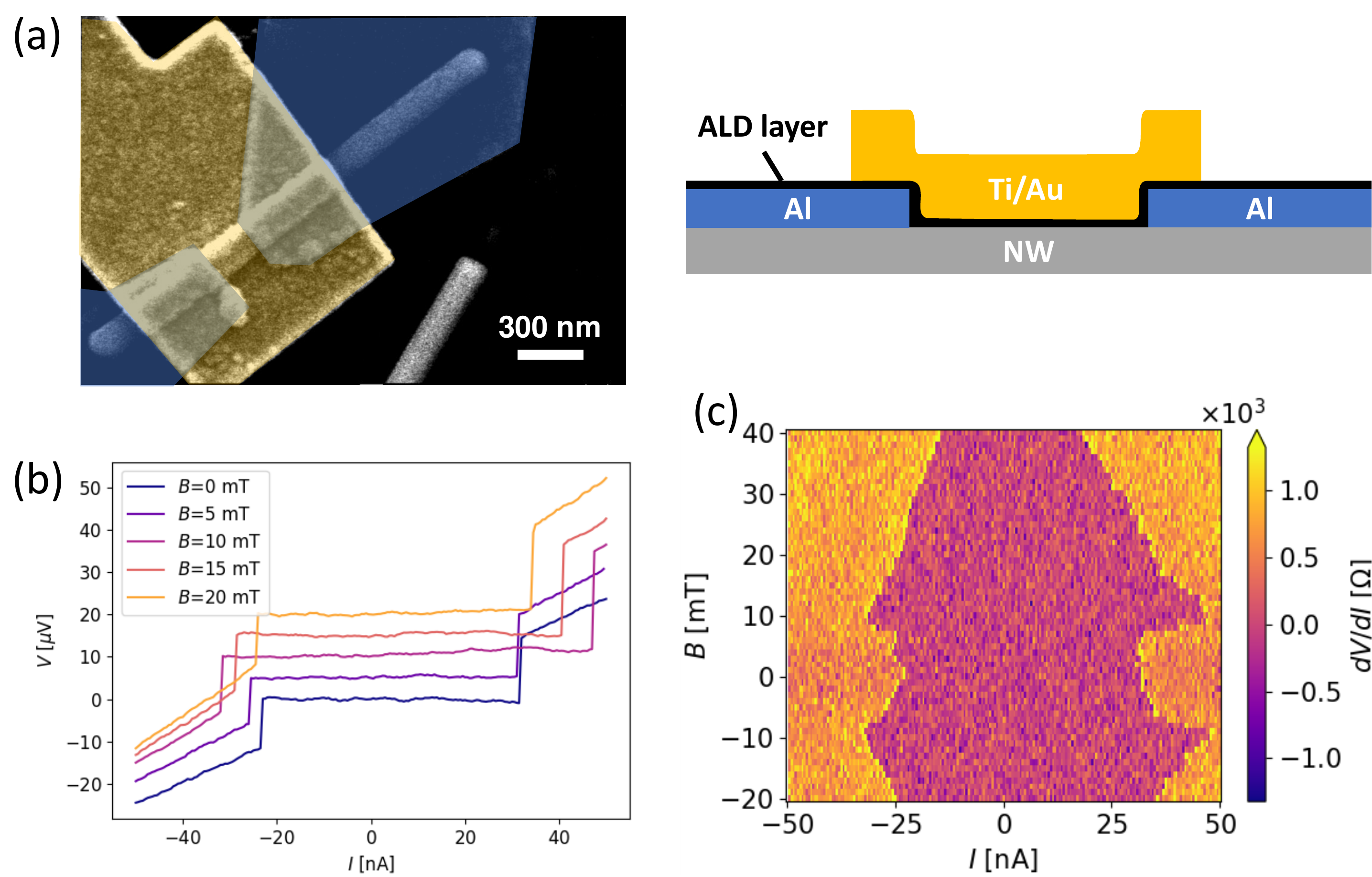}
	\caption{ (a) Left: SEM image of an InAs single NW Josephson junction device with a top gate electrode (yellow). 
		The junction separation between the two Al (blue) electrodes was approximately \SI{200}{nm}.
		The NW looks thicker than \SI{80}{nm} because of the $\mathrm{Al_2O_3}$ layer.
		Right: Schematic view of cross section along NW. 
		(b) Examples of $V$ vs. $I$ measured for various magnetic fields.
		Each curve was offset by \SI{5}{\micro V}.
		(c) $dV/dI$ as a function of $I$ and $B$. $\Isw$ had a maximum at $B = \SI{10}{mT}$.
		\label{Fig.device}
	} 
\end{figure}

\subsection*{Results}
In this study, a Josephson junction was fabricated on an InAs single NW placed on a Si substrate. 
A scanning electron microscopy (SEM) image of the complete device and a schematic picture of cross section along the NW are shown in Fig.~\ref{Fig.device}(a). 
We used two superconductor Al electrodes that were separated by approximately \SI{200}{nm}. 
The carrier density of the NW was controlled using a top gate electrode. 
The voltage $V$ across the junction as a function of the current $I$ measured under various magnetic fields for a bath temperature of $T=\SI{37}{mK}$ is shown in Fig.~\ref{Fig.device}(b).
Figure~\ref{Fig.device}(c) shows the differential resistance $dV/dI$ as a function of the bias current $I$ and out-of-plane magnetic field $B$ for the device at $V_g = \SI{0}{V}$.

The boundary identified by the color change gives the magnitude of the SC and switching current $\Isw$. 
The $\Isw$ at $B=\SI{0}{mT}$ was \SI{30}{nA} and gradually increased as $B$ increased to \SI{10}{mT}, where it reached a maximum.
We denoted this maximum point as $B^*$.
$\Isw$ then decreased and vanished at $B = \SI{60}{mT}$. 
This result is similar to the previous report~\cite{tiira2017magnetically}, 
including in terms of the magnitude of the enhancement. 
It should be noted that $dV/dI$ in the SC region remains zero at all measured $B$. 
This is a significant difference from the in-plane magnetic field case in Ref.~\cite{kamata2018anomalous}, because the enhancement with the in-plane field is derived from partial breakdown of SC due to difference in thickness, which ends up as the formation of low-pass filters, causing the finite $dV/dI$ at $|B|>|B^*|$. 

\begin{figure}[t]
  \centering
     \includegraphics[width=1.0\linewidth]{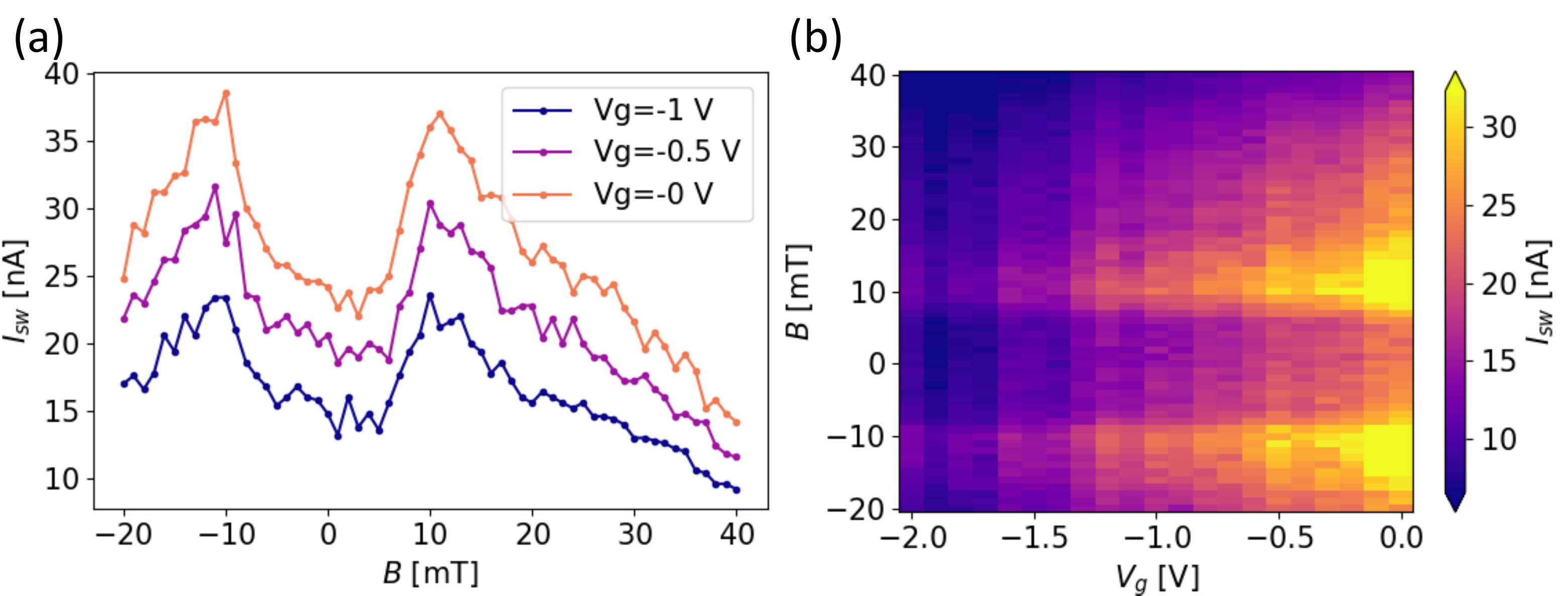}
  
  \caption{ (a) $\Isw$ as a function of $B$ at different $V_{g}$. 
  (b) $\Isw$ as a function of $B$ and $V_{g}$. 
  $B^*$ is found to be independent of $V_g$.
  \label{Fig.Isw_vs_B}} 
\end{figure}

To investigate whether the enhancement originated from the NW or superconducting metals, 
we varied the electron density of the NW by $V_g$. 
Figures~\ref{Fig.Isw_vs_B}(a) and (b) show $\Isw$ as a function of $B$ and $V_g$.
Note that the junction can be completely depleted at $V_g = \SI{-5.1}{V}$ (See S.M.~$\S$1). 
The maximum point $B^*$ remained constant, whereas $\Isw$ changed with varying the electron density of the NW with $V_g$. 
This result indicates that the enhancement did not originate from the NW between the two superconducting electrodes but the superconducting metals or NW beneath the superconducting metals.
If the enhancement originated from the MF contributions, $B^*$ would change as $V_g$ changes because the $B$ corresponding to the topological transition depends on the Fermi energy of the NW. 
Therefore, the observed enhancement of the SC cannot be attributed to the appearance of the TSC phase. 
Because $B^*$ did not depend on $V_g$, the enhancement must have been caused by magnetic field-induced phenomena generated in the superconducting electrodes rather than in the NW.

\begin{figure*}[t]
	\centering
	\includegraphics[width=\linewidth]{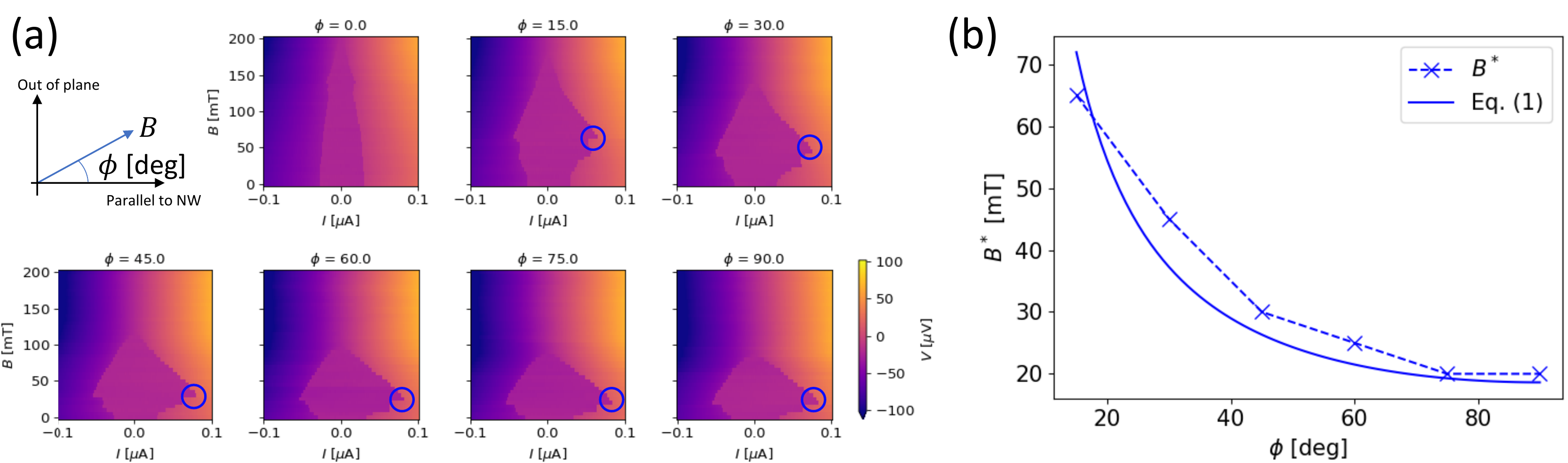}
	\caption{(a) Bias voltage $V$ as a function of $B$ and $I$.
		The blue circles indicate $B^*$.
		(b) Magnetic field $B^*$ vs. $\phi$ (blue crosses). 
		The blue solid line is the fitting result of the $B^*$ vs. $\phi$ data points to Eq.~\eqref{Eq.Bc}.
		The fitting parameter is given as $A=\SI{18.6\pm 1.1}{mT}$.
		\label{Fig.Bstar}
	} 
\end{figure*}

We study $dV/dI$ dependence on $I$ and $B$ by changing the direction of the magnetic field, as shown in Fig.~\ref{Fig.Bstar}(a). 
The magnetic field was applied at an out-of-plane angle $\phi$ measured from the plane. 
We observed an increase in the critical field as the applied $B$ tilts from the out-of-plane ($\phi = \ang{90}$) to the in-plane ($\phi = \ang{0}$) direction.
In Fig.~\ref{Fig.Bstar}(a), the enhancement peak points $B^*$ are highlighted with blue circles.

Here, we fit $B^*$ (crosses in Fig.~\ref{Fig.Bstar}(b)) as a function of the angle $\phi$ using the following formula:
\begin{align}
	B^* &= \frac{A}{\sin\phi} \label{Eq.Bc},
\end{align}
where $A$ corresponds to the out-of-plane component of $B^*$.
The solid line in Fig.~\ref{Fig.Bstar}(b) shows the calculated magnetic field of Eq.~\eqref{Eq.Bc} compared to the experimentally obtained $B^*$.
The good agreement with the experiment indicates that only the out-of-plane magnetic field component determines $B^*$.

\begin{figure*}[t]
	\centering
	\includegraphics[width=1.0\linewidth]{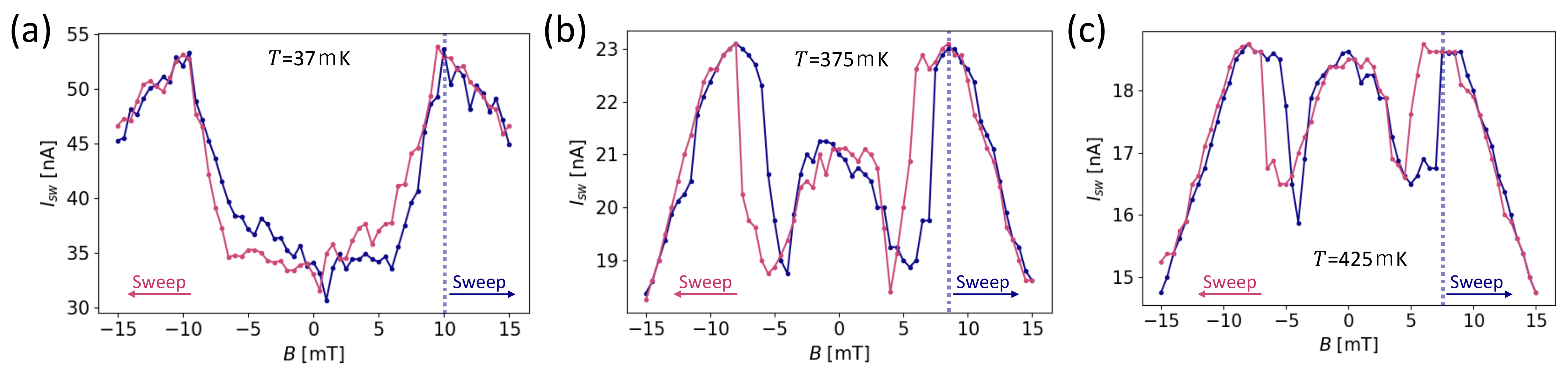}
	\caption{ 
		$\Isw$ vs. $B$ at (a) $T = \SI{37}{mK}$, (b) \SI{375}{mK}, (c) and \SI{425}{mK}. 
		The blue (purple) lines represent the results for the upward (downward) sweep. 
		The hysteresis is visible for $|B|<|B^*|$.
		$B^*$ determined in the upward sweep is shown as dashed lines.
		\label{Fig.Isw_vs_B_Tdep}
	} 
\end{figure*}

Finally, we investigated the $B$ dependence of $\Isw$ when $B$ was swept in different directions. 
Figures~\ref{Fig.Isw_vs_B_Tdep} (a), (b) and (c) show $\Isw$ as a function of $B$ at $T = \SI{37}{mK}$, \SI{375}{mK}, and \SI{425}{mK}, respectively.
The blue (purple) lines represent the upward (downward) sweep.
Here, we found a clear hysteresis of $\Isw$ depending on the $B$ sweep direction, as shown in Fig.~\ref{Fig.Isw_vs_B_Tdep}. 
The hysteresis was apparent for $|B|<|B^*|$ while it did not appear for $|B|>|B^*|$. 
Furthermore, in Fig.~\ref{Fig.Isw_vs_B_Tdep} (a), $\Isw$ in $-B^*<B<\SI{0}{mT}$ is larger than that in $\SI{0}{mT}<B<B^*$ in the sweep from negative to positive and vice versa.
In Figs.~\ref{Fig.Isw_vs_B_Tdep} (b) and (c), the hysteresis appears between $\pm B^*$ and dips around $\SI{\pm5}{mT}$ and has the same sweep direction dependence.
Note that the out-of-plane $B$ dependence of $\Isw$ in Ref.~\cite{tiira2017magnetically} is also asymmetric for $B^*>B>\SI{0}{mT}$ and $-B^*<B<\SI{0}{mT}$, suggesting hysteresis.

\subsection*{Discussion}
We attribute the observed enhancement to quasiparticle trapping by superconducting vortices.
The vortices penetrate a superconductor when a magnetic field applied to the superconductor exceeds the critical field $B_{c1}$. 
Note that supercurrent enhancement has been observed in similar nanowire systems~\cite{Rogachev2006,Murani2019}, but these results do not show hysteresis of $B$ field sweep, meaning the origins of the enhancement are different from ours.
The superconducting pair potential is broken at the vortex cores, and they act as trapping potentials for quasiparticles. 
These quasiparticles corresponds to the excited states in the superconductors. 
Therefore, trapping at the vortex cores makes the thermally excited quasiparticles relax to the bound states in the cores whose energies are lower. 
In the relaxation process, the thermally excited states transfer their energies to the environment (heat bath) as phonons (See S.M.~$\S$4). 
This type of quasiparticle trapping can improve the superconducting device quality, as observed in electron turnstile devices~\cite{Pekola2013,Taupin2016,Nakamura2017}, because the trapping effectively lowers the electron temperature.
This effect has been applied to the design of superconducting qubits, for example, forming vortices in the outer region so that the system of interest is cooled down~\cite{song2009reducing,Peltonen2011,Nsanzineza2014,wang2014measurement}.
The switching current $\Isw$ is affected by thermally excited quasiparticles, depending on the electron temperature. 
Therefore, the observed SC enhancement can be attributed to electron cooling due to quasiparticle trapping.

In this scenario, the observed hysteresis is also reasonable. 
$B^*$ is the point that vortices enters the system, and they always exist at higher $B$ (where $B<B_c$).
The hysteresis appears, for example, when we sweep downward from $B>B^*$.
Here, some magnetic fluxes remain in the system due to pinning effect by impurities or diffraction and provide the cooling effect.
This is consistent with the result that $\Isw$ of downward sweep is larger than one of upward sweep where $0<B<B^*$.
When comparing the results at several temperatures in Fig.~\ref{Fig.Isw_vs_B_Tdep}, it is observed that $B^*$ (dashed lines) decreases with increasing $T$. 
This indicates that $B_{c1}$ becomes smaller as $T$ increases, supporting the electron cooling scenario. 

In addition, the enhancement at $B = B^*$ gradually decreases as $T$ increases. 
This behavior is consistent with the decrease in the electron cooling effect at higher temperatures, because the number of quasiparticles at higher temperatures increases.
The order of the cooling effect can be estimated as \SI{100}{mK} (See Fig.~S5), 
which is comparable to a previous report on normal-metal/insulator/superconductor tunnel junction~\cite{Peltonen2011}. 
We note that the $0$-$\pi$ transition of the junction and magnetic impurities in the NW cannot explain the hysteresis, even if they make $B^*$ remain constant with $V_g$. 
For the same reason, the hysteresis cannot be attributed to the topological transition in the proximitized region.

Our results revealed that the enhancement of the switching current by an out-of-plane magnetic field is non-topological.
However, to realize topological qubits using MFs~\cite{Ivanov2001,Aasen2016}, it is important to reduce the quasiparticle density to protect the information of the qubits from quasiparticle poisoning. 
Even with a higher magnetic field, if the material is a type-II superconductor or thin film in which vortices can penetrate, the device can be designed to trap quasiparticles effectively so that the system of interest is cooled.
Effectiveness of quasiparticle trapping is known to depend strongly on device structure~\cite{Pekola2013}, and therefore further studies for optimal design for cooling will be necessary.
This report shows that quasiparticles in superconducting devices can be reduced by quasiparticle trapping under a finite perpendicular magnetic field, which provides important insights for the design of topological qubit devices in the near future.

\subsection*{Methods}
The InAs NW had a diameter of approximately \SI{80}{nm} 
and was grown on an InAs(111)B substrate by chemical beam epitaxy~\cite{baba2017gate}.
A Josephson junction was fabricated on the NW after transferring it onto a \SI{280}{nm}-thick $\mathrm{SiO_2}$ substrate by standard dry transfer technique with cotton buds. 
Ti/Au markers were fabricated on the substrate in advance, so that we can determine positions of randomly spread NWs.
We made a polymethyl methacrylate pattern for the contact areas using electron beam lithography and performed surface treatment using a $\mathrm{(NH_4)_2S}_x$ solution to remove the native surface oxidised layer. 
Then, the super conducting electrodes were fabricated by depositing Ti/Au (\SI{1}{nm}/\SI{60}{nm}) and lift-off. 
The top gate was fabricated by growth of \SI{20}{nm} thick $\mathrm{Al_2O_3}$ by atomic layer deposition followed by depositing a gate electrodes of Ti/Au (\SI{50}{nm}/\SI{150}{nm})~\cite{baba2017gate,baba2018cooper,kamata2018anomalous,ueda2019dominant,ueda2020evidence}.

\subsection*{Measurement setup}
All measurements were done in a dilution fridge with a standard quasi-4-terminal method. 
The base temperature of the thermal bus was about \SI{35}{mK}. 
The conductance was measured with lock-in amplifiers with an excitation voltage of \SI{10}{\micro V}. 
For the SC measurements, DC voltages across the device were measured with a constant current bias.

For the magnetic field dependent measurements, we swept the field at rate of \SI{0.1}{T/min} and then wait \SI{15}{s} before sweep of bias current.

\subsection*{Acknowledgements}
We thank Prof. S. Jeppesen for the material growth and thank Dr. Shuji Nakamura in AIST for fruitful discussion about thermalization.
This work was partially supported by a Grant-in-Aid for Scientific Research (S) (Grant No. JP26220710), JST PRESTO (grant no. JPMJPR18L8), the JSPS Program for Leading Graduate Schools (MERIT) from JSPS, the ImPACT Program of Council for Science, Technology, and Innovation (Cabinet Office, Government of Japan), Advanced Technology Institute Research Grants, Ozawa-Yoshikawa Memorial Electronics Research Foundation, the Ministry of Science and Technology of China (MOST) through the National Key Research and Development Program of China (Grant Nos.
2016YFA0300601, 2017YFA0303304) and the National Natural Science Foundation of China (Grant Nos. 92165208, 11874071, 91221202, 91421303).

\subsection*{Author contributions}
\begin{acknowledgments}
K.U. and S.M. designed the device;Y.S., K.U., Y.T., H.K., S.M., S.T. joined discussions and their previous results inspired the design; K.U. followed their fabrication process; K.L., L.S., H.Q.X. provided the nanowires used in the experiment; Y.S. performed experiments with input from K.U., S.M. and S.T.; Y.S. and K.U. wrote the manuscript, with input from all authors; S.M. and S.T initiated the project.
\end{acknowledgments}

\end{document}